\documentclass{article}

\usepackage{latexsym}
\usepackage[cp1251]{inputenc}
\usepackage[english,russian]{babel}
\usepackage{graphicx}
\usepackage{amsfonts, amsmath,amssymb}

\renewenvironment{abstract}
 {\begin{center} {\ } \end{center} \quotation}
 {\endquotation}

\newcommand\e{{\mathrm{e}}}

\newcommand\be{\begin{equation}}
\newcommand\ee{\end{equation}}
\newcommand\bee{\begin{eqnarray}}
\newcommand\eee{\end{eqnarray}}

\begin{document}
 \title{On the theory of ideal Bose-gas at a finite particle number
 }{}%
\author{{\it
 A.I. Bugrij\,${}^{1}$\footnote{E-mail: abugrij@bitp.kiev.ua},\addtocounter{footnote}{5}
  V.M. Loktev$^{1,2}$\footnote{E-mail: vloktev@bitp.kiev.ua} }\\
{\small\it ${}^{1}$Bogolyubov Institute for Theoretical Physics of the NAS of Ukraine, }\\
 {\small\it 14-b, Metrolohichna Str., Kyiv 03143, Ukraine}\\
 {\small\it ${}^{2}$National Technical University of  Ukraine ``Igor Sikorsky Kyiv Politechnical Institute`` }\\
 {\small\it 37, Peremogy Ave., Kyiv 03056, Ukraine}}
 \date{}
 \maketitle

The ideal Bose-gas with finite number $N$ of particles is
investigated. The exact expressions for the partition functions and
occupation numbers in the grand canonical, canonical and
microcanonical ensembles are found. The asymp\-totic expressions (in
the case $N\gg1$) for the partition functions and occupation numbers
in the canonical and microcanonical ensembles are
evaluated. It is shown that the chemical potential $\mu$ of the
ideal Bose-gas can lie in the range $-\infty<\mu<\infty$ oppositely
to the widely adopted opinion that the value of this potential is negative.



\section{Introduction}

The theory of ideal Bose-gas is referred to the old and rather well studied field of
statistical physics of macroscopic systems [1-4]. However, the real experiments aimed at
its verification are performed with systems consisting of a finite number $N$ of the particles. 
For example, in the experiments with atomic Bose-gases [5-8] whose results are usually interpreted 
as the experimental confirmation of the phenomenon of Bose-condensation, the number of
particles is at most $\sim10^{4}$, or does not attain the Avogadro number. At the same time, 
no corresponding calculations for the ideal Bose-gas consisting of a finite number of 
particles are available in the literature. Therefore, the goal of the our paper is to fill this gap. 
Moreover, we will calculate the observable quantities for all basic statistical ensembles, namely, 
the grand canonical (GCE), canonical (CE), and microcanonical (MCE) ones.

\section{Ideal Bose-gas}

A quantum particle located in the vessel with impermeable walls
with volume $V$ has the discrete energy spectrum $\varepsilon_{k}$, where the index
$k$ runs the values $0, 1, \ldots, \infty,$ as the energy $\varepsilon_{k}$ of the state increases.

Consider the system consisting of $N$ bosons non-interacting with one another
that are placed into the vessel. The system is characterized by a configuration
$[n]$, i.e., by a set of the occupation numbers $\mathbf{n}_{k}$.
Each of them indicates the number of particles in the $k$-th
state with the energy $\varepsilon_{k}$. It is obvious that
$\mathbf{n}_{k}$ \emph{cannot exceed} the number $N$ of particles in the
system. In this case, their total number and total energy of the system
are given by the equalities:

 \be\label{eq1}
 N=\sum_{k=0}^{\infty}{\mathbf{n}}_{k},\quad
 E=\sum_{k=0}^{\infty}{\mathbf{n}}_{k}\varepsilon_{k}.\ee
 
 Let us place the vessel into a thermostat with the temperature $T$. Under the action of
 thermal fluctuations on the vessel walls, the configuration
 $[n]$ will vary (stochastically) with the time. As is known, from experiments one obtain 
 the time-averaged values of the observable quantities. In statistical physics for their calculations 
 the ergodic hypothesis is as a rule accepted. According to it, the mean over the time coincides with 
 the mean over the ensemble.

 By definition, the ensemble is a collection of systems with all possible configurations
 or, in other words, with different distributions of particles over states.
 The ensembles can be very different. Each specific ensemble
 is determined by the own distribution function $f[n]$, and the most popular among them are, as mentioned above, GCE, CE, and MCE.

\section{Grand canonical ensemble}

The distribution function for GCE takes the form \be\label{eq2}
f[n]=\e^{-\beta(E-\mu N)}=
\prod_{k}\e^{-\mathbf{n}_{k}\beta(\varepsilon_{k}-\mu)}\,,\ee
 where $\beta=(k_{\scriptstyle{B}}T)^{-1}$ stands for the reciprocal temperature,
 and $\mu$ is as above the chemical potential.

Since $f[n]$ (2) is factorized, the partition function of such system can be easily calculated [9]: \bee\label{eq3}Z&=&\sum_{[n]}f[n]=
\sum_{\mathbf{n}_{0}=0}^{N}\e^{-\mathbf{n}_{0}\beta(\varepsilon_{0}-\mu)}
\sum_{\mathbf{n}_{1}=0}^{N}\e^{-\mathbf{n}_{1}\beta(\varepsilon_{1}-\mu)}
\ldots
\sum_{\mathbf{n}_{k}=0}^{N}\e^{-\mathbf{n}_{k}\beta(\varepsilon_{k}-\mu)}\ldots\nonumber\\
&=&
\prod_{k}\frac{1-\e^{-\beta(\varepsilon_{k}-\mu)(N+1)}}{1-\e^{-\beta(\varepsilon_{k}-\mu)}}.\eee
The mean value of the number of occupation can be also determined and takes the form
\be\label{eq4}\overline{n}_{k}=Z^{-1}\sum_{[\mathbf{n}]}\mathbf{n}_{k}f[\mathbf{n}]=n_{k}-
(N+1){m}_{k},\ee where $n_{k}$ and ${m}_{k}$ are defined as follows:
$$n_{k}=\frac{1}{\e^{\beta(\varepsilon_{k}-\mu)}-1},\quad
{m}_{k}=\frac{1}{\e^{\beta(\varepsilon_{k}-\mu)(N+1)}-1}.$$

It is seen that the first term on the right-hand side of (4) is the ordinary Bose-distribution,
and the second one introduces the dependence on the number of particles of the system into the average over the ensemble.
This point is worth to be noted, since the mean number of occupation following from relations 
(3) and (4) coincides formally with that for the parastatistics [3] under the condition $p=N$, 
though it is quite obvious that the order $p$ of the parastatistics and the number  $N$ are different 
physical quantities without any connection with each other. In addition, the derivation of the corresponding 
formula in [3] seems not to be quite proper, because the Stirling factorial formula is used in it for 
the quantities less than 1. Nevertheless, the final result turns out proper and
coincides with formula (4) which is obtained with the help of exact calculations.

It is worth to note that the number  $\overline{n}_{k}$ has a finite value
for $$ -\infty<\mu<\infty,$$ rather than only for $\mu<0$. So, the latter is the assertion which is not absolutely true. 
Indeed, the number of occupation for the lowest state
\be\label{eq6}\overline{n}_{0}=
\frac{1}{\e^{\beta(\varepsilon_{0}-\mu)}-1}-\frac{N+1}{\e^{\beta(\varepsilon_{0}-\mu)(N+1)}-1}.\ee
From whence, we get $\overline{n}_{0}=N$ for $\mu\rightarrow\infty$,
$\overline{n}_{0}={N/2}$ for $\mu\rightarrow\varepsilon_{0}$, and
$\overline{n}_{0}=0$ for $\mu\rightarrow-\infty$. In this case, the parameters $\beta$
and $\mu$ can be presented in terms of $N$ and $E$ by means of the system of
equations (cf. (1)) \be\label{eq7}
N=\sum_{k=0}^{\infty}\overline{n}_{k},\quad
E=\sum_{k=0}^{\infty}\overline{n}_{k}\varepsilon_{k}.\ee

\section{Canonical ensemble}

Unlike the previous case, the distribution function for CE takes the form \be\label{eq8} f_{CE}
[n]=f[n]\delta\biggl(N-\sum_{k}\mathbf{n}_{k}\biggr).\ee The calculation of the
partition function becomes harder due to the presence of the $\delta$-function in (7), 
but the exact analytic calculation can be carried out.

We represent $\delta(N-\sum_{k}\mathbf{n}_{k})$ as the integral: \be\label{eq9} \delta(N-\sum_{k}\mathbf{n}_{k})=\frac
{1}{2\pi}\int\limits_{-\pi}^{\pi}dx\,
\e^{ixN}\prod_{k}\e^{-ix\mathbf{n}_{k}}.\ee Then the partition function of CE takes the form \be\label{eq10} Z_{ {CE}}=\sum_{[n]}f_{
{CE}}[n]=\frac{1}{2\pi} \int\limits_{-\pi}^{\pi} dx\,
\e^{-w(x)}\,\ee where \be\label{eq11} w(x)=-i x N -\sum_{k}
\ln\frac{1-\e^{-[\beta(\varepsilon_{k}-\mu)+ix](N+1)}}{1-\e^{-[\beta(\varepsilon_{k}-\mu)+ix]}},\ee
\be\label{eq12} \overline{n}_{k}^{
{CE}}=\frac{Z_{{CE}}^{-1}}{2\pi}\int\limits_{-\pi}^{\pi} dx\,
\e^{-w(x)} \left(\frac
{1}{\e^{\beta(\varepsilon_{k}-\mu)+ix}-1}-\frac{N+1}
{\e^{[\beta(\varepsilon_{k}-\mu)+ix](N+1)}-1}\right).\ee

 For $N\gg1,$ integrals (9) and (11) can be calculated with the help of the saddle-point method.
Corresponding saddle point is determined from the equation $$\frac{\partial
w(x)}{\partial
x}=0=-i\left(N-\sum_{k}\biggl(\frac{1}{\e^{\beta(\varepsilon_{k}-\mu)+ix}-1}-
\frac{N+1}{\e^{[\beta(\varepsilon_{k}-\mu)+ix](N+1)}-1}\biggr)\right)\,$$
which holds, due to equalities (6), for $x=0$. The second derivative at the
saddle point takes the form \be\label{eq14}
w_{xx}=\left.\frac{\partial^{2}w(x)}{\partial
x^{2}}\right|_{x=0}=\sum_{k}\left(\frac{\e^{\beta(\varepsilon_{k}-\mu)}}
{(\e^{\beta}(\varepsilon_{k}-\mu)-1)^{2}}
-\frac{(N+1)^{2}\e^{\beta(\varepsilon_{k}-\mu)(N+1)}}
{(\e^{\beta(\varepsilon_{k}-\mu)(N+1)}-1)^{2}}\right).\ee As a result, if the condition $N\gg1$ holds, we obtain
\be\label{eq15} Z_{CE}=\frac{Z}{\sqrt{2\pi w_{xx}}}\,.\ee The mean value of the
number of occupation in CE for $N\gg1$ is \be\label{eq16}
\overline{n}_{k}^{CE}=-\beta^{-1}\frac{\partial\ln Z_{CE}}{\partial
\varepsilon_{k}}=\overline{n}_{k}- m_{k}^{CE},\ee where
\be\label{eq17} m_{k}^{CE}=\frac
{1}{2w_{xx}}\left(n_{k}(n_{k}+1)(2n_{k}+1)-(N+1)^{3}{m}_{k}({m}_{k}+1)
(2{m}_{k}+1)\right),\ee and the quantities $n_{k}$ and $m_{k}$ are defined in
(4). The direct verification gives that the numbers of occupation in
the ground state and in the excited ones in CE are, respectively, larger and less, than in GCE.

\section{Microcanonical ensemble}

The distribution function for MCE is easily set and takes the form
\be\label{eq18} f_{
MCE}[n]=f[n]\delta(N-\sum_{k}\mathbf{n}_{k})\delta(E-\sum_{k}\mathbf{n}_{k}\varepsilon_{k})\,.\ee

For the second $\delta$-function, we also use the integral representation analogous to (8), namely:
\be\label{eq19}
\delta\bigl(E-\sum_{k}\mathbf{n}_{k}\varepsilon_{k}\bigr)=\frac{1}{2\pi}
\int\limits_{-\pi}^{\pi}dy
\e^{iyE}\prod_{k}\e^{-iy\mathbf{n}_{k}\varepsilon_{k}}\,,\ee
and can easily sum over the configurations. Then, for the partition function in MCE, we get (cf. (9)) \be\label{eq20} Z_
{MCE}=\sum_{[n]}f_{MCE}[n]=\frac{1}{(2\pi)^{2}}\int\limits_{-\pi}^{\pi}dx
dy \,\e^{-w(x,y)},\ee where \be\label{eq21} w(x,y)=-i N x - i E y
-\sum_{k}\ln\frac{1-\e^{-[\beta(\varepsilon_{k}-\mu)+ix+iy\varepsilon_{k}](N+1)}}
{1-\e^{-[\beta(\varepsilon_{k}-\mu)+ix+iy\varepsilon_{k}]}}\,.\ee
In this case, the mean value of the number of occupation is presented in terms of the integral
\bee\overline{n}_{k}^{MCE}&=&\frac{Z^{-1}_{MCE}}{(2\pi)^{2}}\int\limits_{-\pi}^{\pi}dx
dy \,\e ^{-w(x,y)}\times\nonumber\\
 &\times&\left(
\frac{1}{\e^{\beta(\varepsilon_{k}-\mu)+ix+iy\varepsilon_{k}}-1}-\frac{N+1}{\e^{
[\beta(\varepsilon_{k}-\mu)+ix+iy\varepsilon_{k}](N+1)}-1}\right),\label{eq22}
\eee which can be considered as a generalization of integral (11). In the calculation of the partition function (18), we take
$N\gg1$ and use the saddle-point method whose point has the coordinates
$x=y=0$ . As a result, we obtain
\be\label{eq23}Z_{MCE}=\frac{1}{2\pi}\frac{Z}{\sqrt{d}},\ee where
\be\label{eq24}d=w_{xx} w_{yy}-w_{xy}^{2},\ee $$
w_{xx}=\left.\frac{\partial^{2}w(x,y)}{\partial
x^{2}}\right|_{x=y=0},\quad
w_{xy}=\left.\frac{\partial^{2}w(x,y)}{\partial x
\partial y}\right|_{x=y=0},\quad w_{yy}=\left.\frac{\partial^{2}w(x,y)}{\partial
y^{2}}\right|_{x=y=0},$$

\medskip
 \be\label{eq26}
\overline{n}_{k}^{MCE}=\overline{n}_{k}-m_{k}^{MCE},\ee and
\be\label{eq27} m_{k}^{MCE}=
\frac{1}{2d}\left(\bigl(w_{yy}+\varepsilon_{k}^{2}w_{xx}\bigr)b_{k}-\bigl(2\varepsilon_{k}
w_{xx}-w_{xy}\bigr)\frac{a_{k}}{\beta}\right),\ee

$$a_{k}=n_{k}(n_{k}+1)-(N+1)^{2}{m}_{k}({m}_{k}+1),$$

$$b_{k}=n_{k}(n_{k}+1)(2n_{k}+1)-(N+1)^{3}{m}_{k}({m}_{k}+1)(2{m}_{k}+1).$$

As is seen from formulas (22) and (23), the mean values of the numbers of occupation
of the ground state and of the excited ones in MCE are, respectively, larger and less, than in the above-considered GCE and CE.

\section{Conclusions}

The above results are different from the commonly known ones (see,
e.g., [1-4]) and supplement them.  In our opinion, the cause lies in
that the previous studies (at least, the studies of systems in which
the number of particles (bosons, in this case) is given) neglected
some obvious fact, namely, the physical limitation imposed on the
numbers of occupation $\mathbf{n}_{k}\leq N$, which requires a
special conside\-ration.

Nevertheless, even for a sufficiently large number of bosons that usually corresponds to the
experiments on the Bose--Einstein condensation of cold atomic gases, their
thermodynamic quantities (in particular, the condensate density) depend,
as is shown above, on the number $N$. Therefore, we believe that the presented
results can be useful for the thermodynamics and
statistical physics of Bose-systems with finite numbers of particles.

\medskip
 We are grateful to A.S. Kovalev for his interest in the problem
 and the useful discussions.

\medskip
The work was performed in the frame of the programs 0117U00236 and
0116U003191 (State CPCEC 6541210 and 6541230) and the
Special scientific program 0117U00240 of the
 Department of Physics and Astronomy of the NAS of Ukraine.


\end{document}